# DOUBLE STAGE LYOT CORONAGRAPH WITH THE APODIZED RETICULATED STOP FOR EXTREMELY LARGE TELESCOPE

Natalia Yaitskova*

European Southern Observatory

## ABSTRACT

One of the science drivers for the Extremely Large Telescope (ELT) is imaging and spectroscopy of exo-solar planets located as close as 20mas to their parent star [1]. The application requires a well thought-out design of the high contrast imaging instrumentation. Several working coronagraphic concepts have already been developed for the monolithic telescope with the diameter up to 8 meter. Nevertheless the conclusions made about the performance of these systems cannot be applied directly to the telescope of the diameter 30-100m. The existing schemes are needed to be reconsidered taking into account the specific characteristics of a segmented surface. We start this work with the classical system – Lyot coronagraph. We show that while the increase in telescope diameter is an advantage for the high contrast range science, the segmentation sets a limit on the performance of the coronagraph. Diffraction from intersegment gaps sets a floor to the achievable extinction of the starlight. Masking out the bright segment gaps in the Lyot plane although helps increasing the contrast, does not solve completely the problem: the high spatial frequency component of the diffractive light remains. We suggest using the Lyot stop which acts on the light within gaps in order to produce the uniform illumination in the Lyot plane. We show that for the diffraction limit regime and a perfect phasing this type of coronagraph achieves a sufficient star light extinction.

## 1. INTRODUCTION

The term "high contrast imaging" (HCI) is referred to a large number of optical techniques combined by one common aim. This aim is the detection and imaging of faint objects (planets, planetary disks, companions etc.) on the intensive background formed by scattered and diffracted light from the parent star. Generally speaking there are two characteristic inputs of this task: first is that the background is much brighter than the object (typically the difference is 4-5 orders of magnitude), and second, that the faint object is located relatively close to the star (as close as 20 mas angular separation). The purpose of coronagraph as a part of HCI is to suppress the star light in the area of the expected planet location with minimal suppression of the planet light.

All planets search projects are of two kinds. Thus one speaks about the "telescopes for coronagraph (nulling interferometer)" and designs the telescope (configuration of telescopes) in the way the most suitable for HCI (DARWIN[2], TPF[3]). Or one speaks about "coronagraph for the telescope" and designs the HCI system the most compatible with the telescope (VLT-PF[4], JWST[5]). In a case of ELT[6,7] we are in the second situation. In order to make a choice among a huge variety of HCI techniques and methods one has to have a very clear definition of the imaging system itself and take into account all telescope particularities. The first order task inputs here are: giant pupil of the diameter 30m-100m with a relatively large central obscuration and segmentation. The segments phasing errors must be also considered at the level of coronagraph design. The high precision wavefront control and speckle subtraction are necessary conditions for HCI, nevertheless they are external to the coronagraph itself: adaptive optics is applied before and speckle subtraction after the star extinction module.

Based on the prime input of our task – telescope diameter, – the first logical candidate to be considered for ELT is classical Lyot coronagraph[8] with the absorbing focal mask. This type of the coronagraph we present in the paper, showing its specifics with respect to the segmentation and also the possible way to overcome the segmentation problem.

---

* nyaitsko@eso.org, Karl Schwarzschildstr. 2 D-85748 Garching bei Muenchen, Germany, phone +49 89 3200 6581

## 2. LYOT CORONAGRAPHY

### 2.1 Exposition

Classical Lyot coronagraph is barely efficient for the telescope of diameter from 2 to 8m; it may get a new life for ELT. This scheme (Figure 1) includes the absorbing mask placed in the focal plane to block the core image of the star and the diaphragm in the next pupil plane to cut the remained light diffracted outside the pupil.

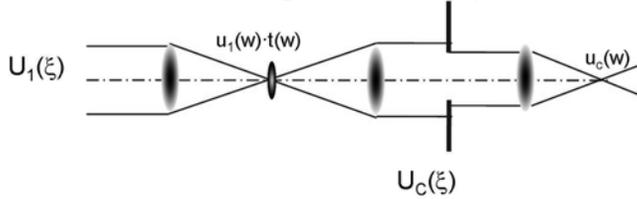

**Figure 1. Principle scheme of Lyot coronagraph with notation of the complex fields in different coronagraphic planes**

Vectors $\boldsymbol{\xi}$ and $\boldsymbol{w}$ were chosen for pupil and image planes correspondingly. We denote by $U_1(\boldsymbol{\xi})$ a complex amplitude, measured in the pupil plane. In the next focal plane its Fourier transform $u_1(\boldsymbol{w})$ is multiplied by the function $t(\boldsymbol{w})$, describing the shape of the absorbing mask. The inverse Fourier transform of that product is the complex amplitude in the next pupil plane (Lyot plane) - $U_C(\boldsymbol{\xi})$. A diaphragm with diameter a bit less than a pupil size cuts the remaining light. In idealistic case this system totally suppresses the deterministic part of the star PSF ("perfect coronagraph"). In reality there is some leak of light, even in diffraction limit case, which contributes to the residual after coronagraph star image, $u_c(\boldsymbol{w})$. The intensity and the shape of the residual $PSF_C=|u_c(\boldsymbol{w})|^2$ characterizes the performance of the coronagraph and defines the contrast.

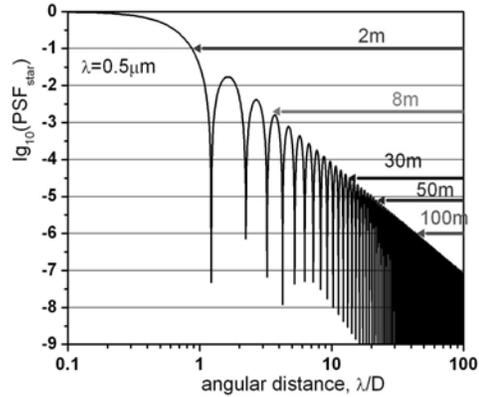

**Figure 2. PSF from the star seen through a circular telescope. The angular distance is measured in units $\lambda/D$. The arrows show the location of the planet at 50mas for different diameter of the telescope and wavelength 0.5μm.**

The efficiency of the Lyot coronagraph depends on the size of the focal mask measured in the number of Airy disks. On the other hand, the mask cannot be very large in order not to block the planet itself. Clearly, the larger telescope diameter the more number of Airy rings is covered by a mask of a given size measures in arc seconds, the more efficient the coronagraph. Figure 2 illustrated this fact. It shows the location of the planet at 50mas with respect to the Airy pattern from the star for 0.5μm wavelength.

For the mathematical convenience it is better to operate not with the function $t(\boldsymbol{w})$, but with the function $m(\boldsymbol{w})=1-t(\boldsymbol{w})$. Following the equations of the Fourier optics it can be shown that the complex amplitude in the Lyot plane before applying a diaphragm is a difference between the initial complex amplitude $U_1(\boldsymbol{\xi})$ and its convolution with the Fourier image of the function $m(\boldsymbol{w})$ :

$$U_C(\boldsymbol{\xi})=U_1(\boldsymbol{\xi})-\frac{1}{\lambda^2}\int U_1(\boldsymbol{\xi}')M(\boldsymbol{\xi}-\boldsymbol{\xi}')\,d^2\xi' \qquad (1)$$

Note, that as function $m(\boldsymbol{w})$ equals unit in the center ($t(0)=0$), its Fourier transform $M(\boldsymbol{\xi})$ has an aria equal to $\lambda^2$ regardless the shape of the mask:

$$\int M(\boldsymbol{\xi})d^2\xi=\lambda^2\cdot \qquad (2)$$

To achieve a full extinction of the star light the function $Uc$ must equal zero within the pupil. It can be achieved by the choice of the mask profile such that $U_1=U_1\otimes M$ within the pupil. From Eq.2 it follows that function $M(\boldsymbol{\xi})$ must be limited within the pupil. Mask with this property is called band limited [9]. Top-hat mask is not band limited. The rule of Fourier transform tells us that the band limited mask is not limited itself, i.e. has infinite wings. The planet may fall in one of the maxima of the wings, which leads to the lost of throughput: part of the planet photons is not absorbed by a coronagraph. Although, if the size of the mask is small or the planet is far from the star, the band-limited masks are efficient. In the following we study a case of a mask with the Gaussian profile, which behaves similar to the band-limited one, but easier for analytics. Note also, that the efficiency of a double stage coronagraph, describes in the section 3, will increase with the band limited mask in comparison with the results presented for the Gaussian one.

## 2.2 Monolithic telescope

To understand the main properties of the Lyot coronagraph we consider first the telescope with the non-segmented aperture diameter $D$, without central obscuration, uniformly illuminated:

$$U_1(\xi) = \begin{cases} U_0, \xi < D/2 \\ 0, \xi > D/2 \end{cases} \tag{3}$$

For the Gaussian mask function $m(w)$ and its Fourier transform are

$$m(w) = \exp\left[-\left(2\sqrt{\ln 2}\, w/a\right)^2\right], \qquad M(\xi) = \frac{b^2 \lambda^2}{\pi} \exp\left[-\left(b\xi\right)^2\right], \tag{4}$$

with $a$ being the full width half maximum of the mask. For the compact notation we assume the approximate values of the coefficient $\pi/2\sqrt{\ln 2} \approx 1.89$, therefore $b = \pi/2\sqrt{\ln 2}\, a/\lambda \approx 1.89\, a/\lambda$. If $a$ is expressed in fraction of Airy disks then:

$$a = s\lambda/D, \quad b \approx 1.89\, s/D \tag{5}$$

Substituting Eqs. 3 and 4 into Eq. 1, and using central symmetric properties of all functions involved, for the complex amplitude in Lyot plane we obtain:

$$U_C(r) = U_1(r) - U_0 \frac{p}{\sqrt{\pi}} \int_{-1}^{1} dt \exp\left[-p^2\left(t - \frac{2r}{D}\right)^2\right] \Phi\left(p\sqrt{1 - t^2}\right). \tag{6}$$

where $r$ is a radial distance from the center of the pupil, $\Phi(x)$ is an error function and $p = bD/2 \approx 0.942\, s$. The example of $U_C$ is shown in Figure 3 with $U_0 = 1$ for $s = 2$, 8 and 16. Function $U_C$ depends only on the distance from the center and parameter $s$. In the center $r = 0$

$$\lg[U_C(0)/U_0] \approx -0.4\, s^2. \tag{7}$$

For $s > 5$ intensity in the middle of Lyot plane can be considered as zero. For the band limited mask it is exact zero everywhere within the pupil except for the border area.

The residual light is concentrated near the border and must be removed by a diaphragm. With an increase of $s$ this area narrows and the diaphragm becomes larger, increasing the throughput for the planet.

To obtain the amplitude spread function (*ATF*) after the coronagraph we must apply the Fourier transform to the function $U_C(\xi)$ taking into account a diaphragm. Because this function is central symmetric, the ASF is also symmetrical and Fourier transform becomes Hankel transform of zero order:

$$u_C(w) = U_0 \int_{0}^{\gamma} 2J_0\left(\frac{\pi D}{\lambda} w\rho\right)\rho d\rho \left\{1 - \frac{p}{\sqrt{\pi}} \int_{-1}^{1} \exp\left[-p^2(t - \rho)^2\right] \Phi\left(p\sqrt{1 - t^2}\right) dt\right\}. \tag{8}$$

where $\gamma$ is linear reduction of the pupil by diaphragm: $\gamma = D_{Lyot}/D$.

As we are interest in a residual contrast we normalize the residual star point spread function by the maximum of the *PSF* before the coronagraph:

$$PSF_C = |u_C(w)|^2 / |u_1(0)|^2. \tag{9}$$

Note that for a give angular distance measured in units $\lambda/D$ the $PSF_c$ is defined only by parameters $s$ and $\gamma$.

The contrast is defined by maxima of the PSF. We denote the curve connecting the maxima by $\Gamma(x)$ (Figure 7). Without coronagraph $\Gamma(x)$ connects the maxima of Airy pattern. For a circular telescope of diameter $D$ and central obscuration ration $\Delta$ this function is given by an approximation:

$$lg[\Gamma(x)] \approx -3lg(x) - 1.08 + 1.23\Delta^{0.8} , \tag{10}$$

where $x > 0$ is an angular distance measured in number of Airy disks: $x = wD/\lambda$. Coefficients 3, 1.08 and 1.23 were found numerically. The central peak ($x=0$) does not lie on the curve.

It can be shown that for angular distances $>10\lambda/D$ the analogous curve can be found for the contrast estimation after the coronagraph. The modification is in an additive parameter $K(s,\gamma) > 0$ which the function of mask size and diaphragm diameter:

$$lg[\Gamma_C(x)] \approx -3lg(x\gamma) - 1.08 + 1.23\Delta_\gamma^{0.8} - K(s,\gamma), \tag{11}$$

where $x$ remains an angular distance measured in $\lambda/D$ units. We included the reduction of the pupil by a diaphragm. For $\Delta \neq 0$ the increase of the central obscuration must be taken into account: $\Delta_\gamma = (\Delta + 1 - \gamma)/\gamma$, $\Delta_\gamma = 0$ if $\Delta = 0$. Example $lg[\Gamma_C(20)]$- contrast at the distance $20\lambda/D$, - is plotted in Figure 4 as a function of $s$ for a set of $\gamma$. This plot shows which combination of mask and stop is required to achieve a given contrast at the distance $20\lambda/D$. For example, suppose that for a given wavelength and a telescope diameter planet star separation equals $8\lambda/D$. Maximum mask size we define in a way to have a lost of planet throughput due to the Gaussian mask not more than 1%. That condition gives $s=10$. From the Figure 4 we obtain that to have a contrast $10^{-10}$ at $20\lambda/D$ for $s=10$ the linear size of the pupil must be reduced by 20%, i.e. $\gamma=0.8$. As it has been mentioned before, the performance (in a sense of the contrast-throughput trade-off) of the coronagraph with the band limited mask is better than that for the Gaussian mask.

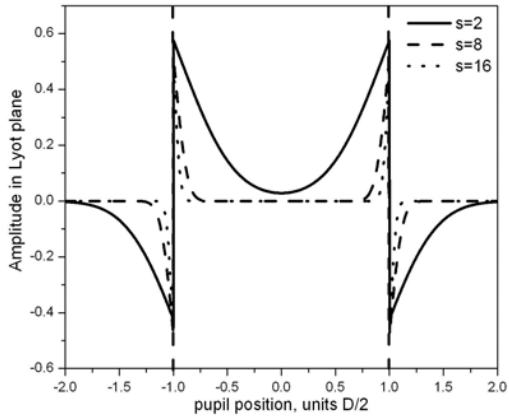

**Figure 3. Amplitude profile in Lyot plane for several sizes of the mask s=2, 8, 16.**

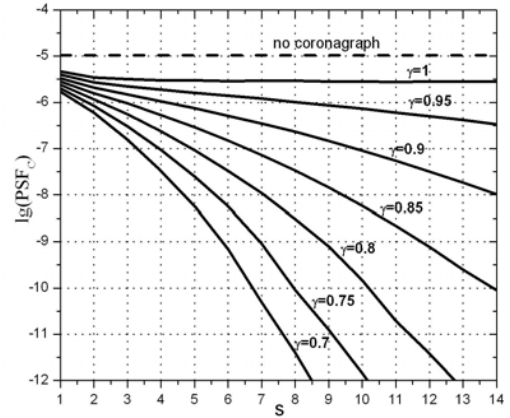

**Figure 4. Contrast (maximum of PSFc) at angular distance 20λ/D for circular telescope without central obscuration as a function of mask size for different size of Lyot stop**

## 2.3 Segmented telescope with the gaps

Segmentation cardinally changes the picture [10]. Function $U_1$ is now a sum of the top-hat functions shifted from each other by the distance equal to segment center-to-center separation:

$$U_{1,seg}(\xi) = U_0 \sum_{j=1}^{N} \Theta_j(\xi - r_j). \tag{12}$$

We denote: $N$ is the total number of segments, $\xi$ is the position vector in the aperture plane, $j$ is the segment index, $r_j$ is the segment center point, and $\Theta_j$ is the segment transmission function:

$$\Theta_j\left(\mathbf{x}-\mathbf{r}_j\right)=\begin{cases}1 & \text{inside the segment aperture}\\ 0 & \text{outside the segment aperture}\end{cases}.\qquad(13)$$

At this stage we consider only effect of the inter-segment gaps, assuming that the telescope is perfectly phased. The questions concerning the residual phasing errors are discussed in section 3.

To calculate the complex amplitude in Lyot plane we have to convolve function $M(\xi)$ with a comb function from Eq.12. It can be shown that if between the size of the mask $s$, telescope diameter $D$, and the segment size $d$ the following condition is held

$$2 < s < 0.7\,D/d\,,\qquad(14)$$

then the convolutions $U_1 \otimes M$ and $U_{1\,seg} \otimes M$ are related as

$$U_{1seg} \otimes M \approx (1-2\omega)(U_1 \otimes M).\qquad(15)$$

We introduced here the parameter $\omega$ - ratio between an averaged size of the gap $g$ to the segment size $d$: $\omega=g/d\ll1$. The easiest way to prove Eq.15 is to consider one dimensional telescope continuing to the two dimensional case of the telescope with a square geometry. The result is the same for any segmentation geometry:

$$U_{C\,seg}(r)/U_0 \approx U_{1seg} - (1-2\omega)\frac{p}{\sqrt{\pi}}\int_{-1}^{1}\exp\left[-p^2\left(t-\frac{2r}{D}\right)^2\right]\Phi\left(p\sqrt{1-t^2}\right)dt\,.\qquad(16)$$

The amplitude in the Lyot plane $U_c$ is shown in Figure 5 ($U_0$=1). The case with the gaps ($\omega$=0.017) was calculated by simulations. For comparison, we also plot the analytical profile of $U_c$ for the same $s$ but without gaps calculated from Eq.6.

For $s>5$ the non-uniformity of $U_C$ within the diaphragm can be neglected so that $U_{1seg} \otimes M \approx (1-2\omega)U_0$ and therefore

$$U_C(\xi)\approx\begin{cases}2\omega U_0 & \text{inside the segment aperture}\\ -(1-2\omega)U_0 & \text{in the gap}\end{cases}.\qquad(17)$$

For the band limited mask, the equality is exact.

The gaps in the Lyot plane are bright and have negative complex amplitude. The area within the segments is constantly illuminated. Two possible types of Lyot stop are evident. First possibility (Lyot stop type 1) is to keep only the diaphragm without removing the light from the bright gaps. Second possibility (Lyot stop type 2) is to manufacture an absorbing grid, duplicating the segmentation grid, totally masking the light in the gaps. As a gap for a planet is an absorbing area, the second type of Lyot stop does not affect the light from the plant. Even if this type of the stop can be accurately manufactured and precisely aligned, the second type of coronagraph does not give a full extinction due to the remaining "floor" $2\omega U_0$.

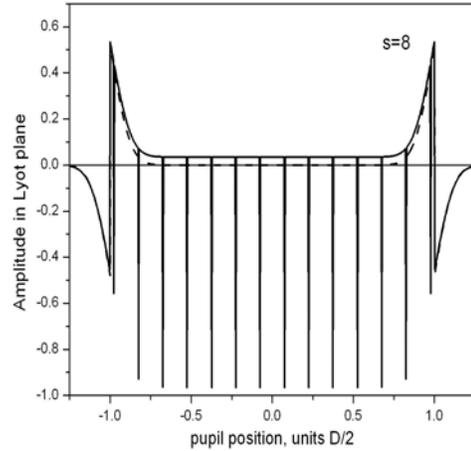

**Figure 5. Amplitude in the Lyot plane for the telescope with (solid) and without (dash) gaps**

To compare the residual *PSFc* for two types of Lyot stop we use the representation of the *PSF* for segmented mirrors developed in [11]. The *PSF* for segmented mirror can be represented as the product of a periodical function of sharp peaks (grid factor - *GF*) and the *PSF* produced by one individual segment (*PSFS*). For the *PSFC* after coronagraph the analogous representation is valid. The application of Lyot stop on $U_C$ reveals in the change of function *PSFs*, while the *GF* remains unchanged. For two types of Lyot stop

$$PSF_C=\begin{cases}GF\cdot PSF_{s,cor}, & \text{type 1}\\ GF\cdot 4\omega^2\cdot PSF_s & \text{type 2}\end{cases}.\qquad(18)$$

All functions are shown in Figure 6. For the first type of Lyot stop $PSF_{s,cor}$ coincides with $PSFs$ in the location of *GF* peaks. Consequently, the coronagraph with stop type 1 does not affect the diffraction peaks in the *PSF* and reduces the central peak. For the second type of Lyot stop the difference with the *PSF* before coronagraph is in the presence of the multiplier $4\omega^2$. Therefore $PSFc = 4\omega^2 PSF$ everywhere including peaks.

Figure 6 cannot serve for the contrast estimation; curves are obtained for the exaggerated gaps to illustrate the principle. Nevertheless, qualitatively the relative behavior of *GF*'s multipliers remains the same for a typical gap size of 5-10mm; and the efficiency of the second scheme remains much higher than for the type 1. The throughput in both schemes is the same.

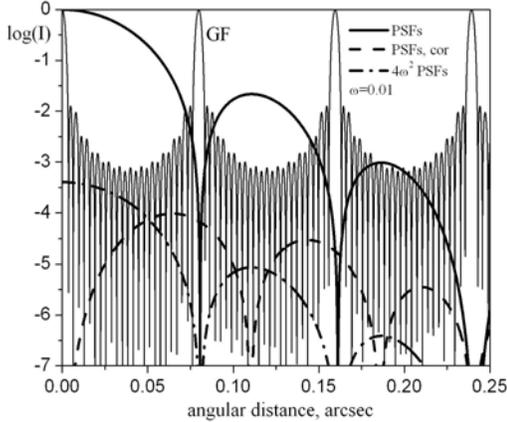

**Figure 6. Formation of the PSF for the segmented mirror before and after coronagraph. Function GF is multiplied by PSFs before, by PSFs,cor or $4\omega^2$PSFs after coronagraph, according to the type of the Lyot stop.**

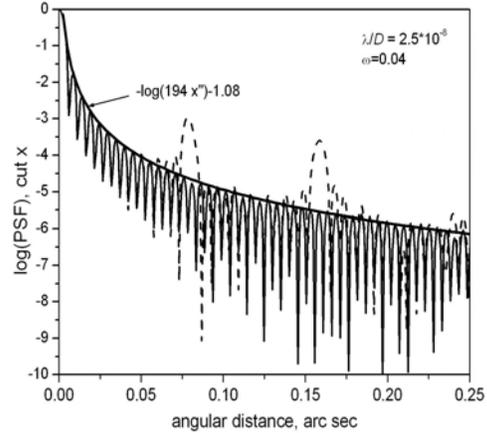

**Figure 7. Normalized PSF for the monolithic (solid) and segmented with the gaps (dash) telescope of the same diameter. Δ=0 The angular distance is shown for λ/D=2.5* 10⁻⁸. The thick line is the analytical curve for the contrast estimation.**

In the following we estimate the efficiency of the second scheme. The pupil after this stop can be considered analogous to the pupil of the segmented telescope with the same gaps and uniformly illuminated with amplitude $2\omega U_0$. This telescope has a characteristic for segmented telescope *PSF* with the satellite peaks. Therefore, for the efficiency estimation we need to calculated the *PSF* from segmented mirror and then to multiply it by a factor $4\omega^2$.

In the expressions for the *PSF* for segmented mirrors [11] the circular outer and inner telescope masks were not taken into account. If the latter is included than the Airy rings start being a dominated pattern of the *PSF*. Suppose that there are no piston-tip-tilt segment errors, but only intersegment gaps are considered. As it is illustrated in Figure 7, everywhere except for the vicinity of the diffractive peaks, the maxima of the normalized *PSF* (related to contrast) from the segmented mirror with the gaps coincides with maxima of the normalized *PSF* from the same mirror without intersegment gaps. The intensity of the peaks can be defined by the Eq.18. The contrast within the area of $\pm 2\lambda/D$ around each diffractive peak can be defined separately. This approach for contrast estimation is applicable as well to the telescope with the central obscuration.

Combining Eq.10 and 18 and taking into account the Strehl ratio due to the gaps equal to $(1-\omega)^2$ we obtain an expression for the contrast *around the peaks* for the second type of Lyot stop:

$$\lg\left[\Gamma_\omega(x)\right] \approx -3\lg(x\gamma) - 1.08 + 1.23\Delta_\gamma^{0.8} + 2\lg(2\omega) . \qquad (19)$$

We approximated $\lg\left[(1-\omega)^2 4\omega^2\right]$ by $2\lg\left[2\omega\right]$.

The intensity of the *diffractive peaks* in PSFc comes from the Eq. 16. For example, the series of A-peaks [11] shown in Figures 6 and 7 in PSF before coronagraph have intensities: $I_{A1} = 0.68\omega^2$, $I_{A2} = 0.17\omega^2$, $I_{A3} = \omega^4$. After the coronagraph their intensity is multiplied by $4\omega^2\gamma^4$, which gives

$$I_{A1,cor} = 2.74\,\omega^4\gamma^4, \; I_{A2,cor} = 0.68\,\omega^4\gamma^4, \; I_{A3,cor} = 4\omega^6\gamma^4, \tag{20}$$

The location of $A$-peaks in $\lambda/D$ units is $x_{An} = n\,2D/\sqrt{3}d$, n=1,2…. Factor $\gamma^4$ appeared because of the normalization $|u_C(w)|^2$ by $|u_1(0)|^2$.

Using Eq. 19 and 20, we plotted in Figure 8 the estimated contrast for two typical size of the gap g=5mm and g=14mm. For size of the segment $d$=1.6m it gives correspondingly ω=0.003 and ω=0.009. We assume that the size of the focal mask is such that γ=0.8 provides a flat light distribution after the diaphragm, so the result is independent on the size of the focal mask.

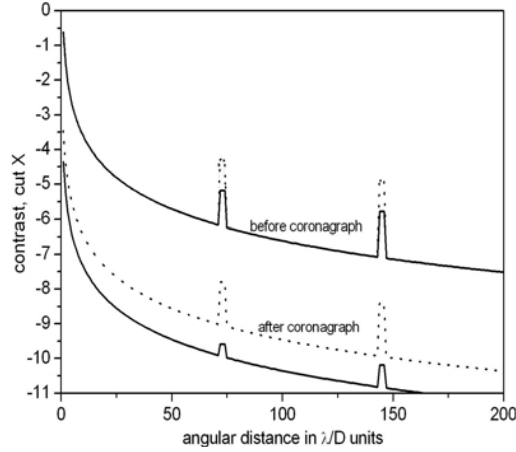

**Figure 8. Estimated contrast for $D/d$=63, Δ=0.3, γ=0.8 and gap size: ω=0.003 (solid) and ω=0.009 (dotted)**

# 3. DOUBLE STAGE SYSTEM

## 3.1 Solving gaps problem
We have seen that the diffraction by gaps sets a floor to extinction achievable with the coronagraph in a classical scheme. The required contrast depends on scientific parameters: planet brightness and angular position, wavelength. Nevertheless, if the scientific requirements are such that the contrast estimated in Figure 8 is not sufficient, the coronagraph can be modified to improve it. Of course this improvement is at the expense of the set up complications. The double stage scheme we present here is a possible modification.

According to Eq. 17, there is proportionality between complex amplitude $U_C$ within the segment pupil and the gap. So far we were removing the component $U_{C,Gap}$ completely by placing the reticulated Lyot stop with zero transmission in the gap, creating again "segmentation" in the Lyot plane. This can be avoided if we change the wave front in the gap area such that $U_{C,Gap} = U_{C,Segment}$, creating a uniform illumination in Lyot plane. For that we have to inverse the phase of the wavefront and attenuate the amplitude by a factor $2\omega/(1-2\omega)$. Complex amplitude after this stop is

$$U_{CL} = U_C(\xi)L(\xi)P_{\gamma 1}(\xi), \tag{21}$$

where $P_{\gamma 1}$ describes a diaphragm with outer diameter $\gamma_1 D$ and, in case of the central obstruction, inner diameter $(\Delta + 1 - \gamma_1)D$. Function $L(\xi)$ describes the amplitude - phase grating:

$$L(\xi) \approx \begin{cases} 1 & \text{inside the segment aperture} \\ 2\omega/(1-2\omega)\,e^{i\pi} & \text{in the gap} \end{cases} . \tag{22}$$

Note that the attenuation factor does not depend on the intensity of the light (star brightness). This modification of the grating does not affect the planet image, because the gaps for the planet are absorbing. Neglecting the non-uniformity of $U_C$ and the remaining effect of the segmentation near the border of the diaphragm for $U_{CL}$ we can write

$$U_{CL}(\xi) \approx \begin{cases} 2\omega U_0, & \xi < D\gamma_I/2 \quad \text{and} \quad \xi > D(\Delta+1-\gamma_I)/2 \\ 0, & \text{elsewere} \end{cases} \tag{23}$$

Again, for the band-limited mask with corresponding choice of the diaphragm, the equality is exact.

Placing a diaphragm in the first we decrease the telescope diameter and hence the throughput. Following the same principle we may suggest to modify $U_c$ on the pupil border and even outside to have a uniform illumination everywhere within the initial pupil boundary. Nevertheless in this paper we do not proceed with this study and consider the case of a diaphragm.

The complex amplitude $U_{CL}$ does not carry anymore the segmentation; and the second stage coronagraph with a simple Lyot stop ideally provides a full extinction. The second stage does not require the reticulated stop, only a diaphragm. The principle scheme is shown in Figure 9.

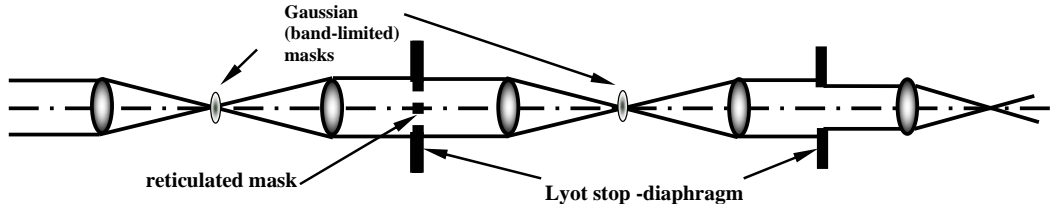

**Figure 9. Principle of a double stage Lyot coronagraph: the reticulated mask inverses the phase and attenuate the intensity; second stage acts like on a monolithic pupil**

### 3.2 Performance

If we assume that the focal masks in the first and the second stages are identical, then with the use of Eq.23 for the complex amplitude in second Lyot plane we obtain:

$$U_{C2} = U_{CL} - (U_{CL} \otimes M) \approx 2\omega U_C(r/\gamma_1, s\gamma_1). \tag{24}$$

where $U_c$ is given by Eq.6. Coordinate and mask size in Eq.6 should be modified to take into account the reduction of the pupil size $\gamma_1$. Function $U_{C2}$ therefore will have all the properties of the function $U_C$. The next Lyot stop diameter $\gamma_2 D$ removes the light on the aperture border ($\gamma_2 < \gamma_1$). For the contrast estimation after the second stage we take into account the non uniformity of $U_C$, which we neglected in section 2.3 and return to the analysis of section 2.2. Applying Hankel transform on $U_{C2}$

$$u_{C2}(w) = 2\omega U_0 \gamma_1^2 \int_0^{\gamma_2/\gamma_1} 2J_0\left(\frac{\pi D\gamma_1}{\lambda}w\rho\right)\rho d\rho \left\{ 1 - \frac{p\gamma_1}{\sqrt{\pi}}\int_{-1}^1 exp\left[-(p\gamma_1)^2(t-\rho)^2\right]\mathfrak{P}\left(p\gamma_1\sqrt{1-t^2}\right)dt \right\}. \tag{25}$$

and normalizing $|u_{C2}(w)|$ by $|u_t(0)|^2$ we obtain a star $PSF_{c2}$ after the second stage. The central obscuration must be included in the low limit of the integral over $\rho$.

Using contrast curve $\Gamma_c(x)$ given by Eq.11 and taking into account the ratio Eq.22 for the contrast curve after the second scheme we obtain

$$\lg[\Gamma_{C2}(x)] \approx -3\lg(x\gamma_1) - 1.08 + 1.23\Delta_{\gamma2}^{0.8} - K(s\gamma_1, \gamma_2/\gamma_1) + 2\lg(2\omega) + 4\lg(\gamma_1), \tag{26}$$

where $\Delta_{\gamma2} = (\Delta+1-\gamma_2)D$. Figure 10 shows the contrast at angular distance $20\lambda/D$ for $\gamma_1 = 0.9$ and different $\gamma_2$. The latter defines the final throughput.

For the classical scheme considered in section 2, the only sensitive to the chromaticity effect is the size of the mask $s$. Nevertheless as we consider the case of large s, the chromatic effect is negligible. On the contrary, for the double stage coronagraph the $\pi$-shift in the first Lyot plane makes a new set-up wavelength dependant. If the mask is designed for wavelength $\lambda_0$, than for $\lambda=\lambda_0+\Delta\lambda$ the complex amplitude in a first Lyot plane after the stop becomes

$$U_{CL}(\xi) \approx \begin{cases} 2\omega U_0, & \text{in the segment aperture} \\ 2\omega U_0 \exp(-i\varphi_\lambda), & \text{in the gap} \\ 0, & \xi > D\gamma_1/2 \quad \text{or} \quad \xi < D(\Delta+1-\gamma_1)/2 \end{cases} \quad (27)$$

where $\varphi_\lambda = \pi\Delta\lambda/\lambda_0$ and we assume that $\Delta\lambda << \lambda_0$. Using representation Eq.21 and a diaphragm function $P_{\gamma 1}$, we re-write

$$U_{CL}(\xi) \approx 2\omega \exp(-i\varphi_\lambda)U_0 P_{\gamma 1} + 2\omega[1-\exp(-i\varphi_\lambda)]U_{1,\text{seg}}(\xi)P_{\gamma 1} = \kappa_1 U_0 P_{\gamma 1} + \kappa_2 U_{1,\text{seg}}(\xi)P_{\gamma 1} \quad (28)$$

where $\kappa_1 = 2\omega \exp(-i\varphi_\lambda)$ and $\kappa_2 = 2\omega[1-\exp(-i\varphi_\lambda)]$. We separated $U_{CL}(\xi)$ into two components: "monolithic" $\kappa_1 U_0 P_{\gamma 1}$ and "segmented" $\kappa_2 U_{1,\text{seg}}(\xi)P_{\gamma 1}$. Consequently, the residual complex amplitude in a focal plane $u_{C2}$ after the second stage coronagraph might be presented as a sum of two components. The first component is proportional to the residual ASF after coronagraph applied on monolithic pupil. The second component is proportional to the residual ASF after coronagraph applied on segmented pupil with the first type of Lyot stop (see text following Eq.18).

The chromatic effect reveals in lost of contrast and appearance of the diffractive peaks determined by a second "segmented" component of $u_{C2}$. The detail analysis for contrast we plan for a future work. Nevertheless now we can make estimation of the peaks intensity. As the first type of the Lyot stop does not alter the peaks, their intensity can be defined directly from Eq.27. For example for A-peaks

$$I_{A1,cor} = 0.68\omega^2(2\omega)^2\gamma_1^4 2[1-\cos(\varphi_\lambda)], \quad I_{A2,cor} = 0.171\omega^2(2\omega)^2\gamma_1^4 2[1-\cos(\varphi_\lambda)], \quad I_{A3,cor} = \omega^4(2\omega)^2\gamma_1^4 2[1-\cos(\varphi_\lambda)], \quad (29)$$

The wavelength dependence is contained in the coefficient $|\kappa_2|^2 = 2[1-\cos(\varphi_\lambda)]$. Figure 11 shows the contrast in the location of the $A_1$ peaks. For the considered example with $D/d$=63 the $A_1$ peaks are located at the distance $73\lambda/D$ from the center. For the monochromatic regime ($\Delta\lambda/\lambda$=0) the contrast at this point is defined by Eq.26. With the increase of $\Delta\lambda/\lambda$ the contribution of the component $I_{A1,cor}$ increases. For comparison on the same graph we plotted the intensity of the $A_1$ peaks after the one stage coronagraph with Lyot stop of the second type. For the large bandwidth the performance of the double stage coronagraph is comparable with the performance of its simplified ancestor.

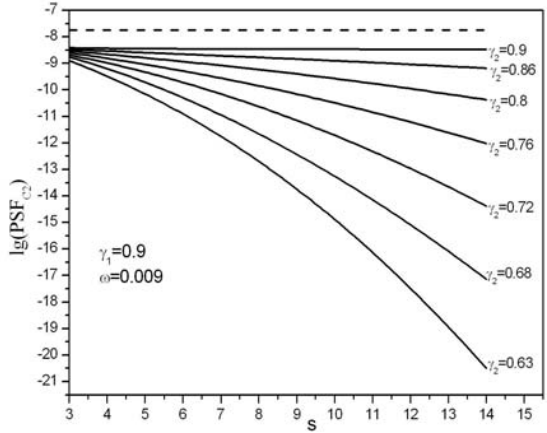

**Figure 10. Contrast at $20\lambda/D$ after the double stage coronagraph as a function of mask size for different size of Lyot stop. Diaphragms in first and second stages $\gamma_1$ and $\gamma_2$. Dash line is a maximal contrast without a second stage. $\Delta$=0.3**

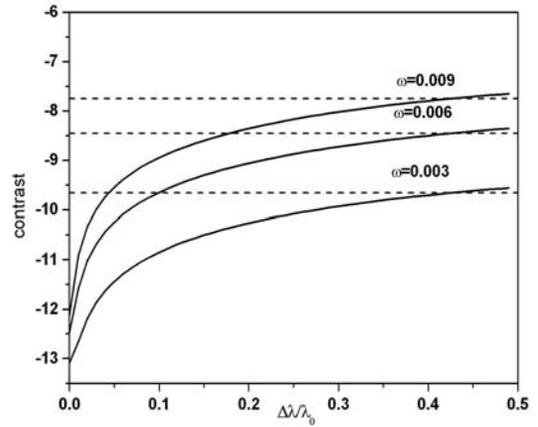

**Figure 11. Contrast at the location of the first diffractive peaks $A_1$ as a function of relative bandwidth for different gap size, $\omega$. $\gamma_1$=0.9, $\gamma_2$=0.72. Dash lines – intensity of the peaks after one stage coronagraph with the second type of the stop.**

### 3.3 Coping with phasing errors

The principle of a double stage coronagraph can be utilizes to reduce the residual phasing error and release the requirement on the residual phasing *rms*. L. Jolissaint and J.-F. Lavigne have demonstrated [12] that the adaptive optics corrects any static aberrations up to spatial frequency system's bandwidth $1/2d_a$, where $d_a$, is separation between actuators. If the wavefront on input of the adaptive optics system consists of segments static aberrations (including piston-tilt-tilt), the in the residual wavefront the phase error remains only in the vicinity $\pm d_a$ around the segments border (Figure 12). Masking the borders with an absorbing grid, we create the segmented telescope with the large gap ($\sim d_a$), although with the decreased phase error. Applying the double Lyot coronagraph on the masked mirror may increase the contrast in comparison with a simple coronagraph applied on the non-masked pupil. As the gaps after the reticulated mask are large and regular, the implementation and alignment of the phase shift attenuating Lyot stop becomes easier. On the other hand the increase of the gap size makes the double Lyot coronagraph more sensitive to the chromatic effects, as it was shown before. Besides, this scheme decreases the throughput. The efficiency of such a system with respect to segmentation error and the contrast-throughput trade-off is the subject of a future study.

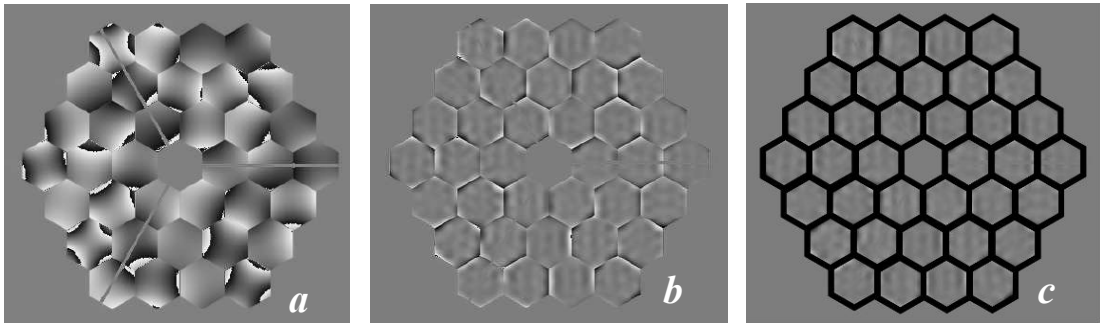

**Figure 12. (a,b) Wavefront from segmented mirror before and after adaptive optics correction. Initial wavefront parameters: 30 nm piston, 0.015" tilt, 30 nm astigmatism. Adaptive optics parameters: $d_a$=0.14$d$, aliasing included. Courtesy Laurent Jolissaint. (c) With the reticulated mask applied.**

## 4. CONCLUSION

Classical Lyot coronagraph for ELT benefits from the size of the focal mask. If the planet located at $4\lambda/D$ for 8m telescope, then for the same wavelength it will be located at $15\lambda/D$ for 30m telescope and $50\lambda/D$ for a 100m telescope.

The concept of segmentation increases the number of effects affecting the performance of the coronagraph. Among them there are the intersegment gaps which set a limit for the achievable star light extinction even for a very huge focal mask. This problem can be solved at the expense of the complication of the system. One of the solutions is the double stage coronagraph with the complex stop in a first stage. This concept allows increasing of contrast by 3-4 orders of magnitude. Typically, the one stage coronagraph applied on segmented telescope with 10mm gap and 1.6m segment gives a contrast $\sim 10^{-8}$. The second stage will decrease this number to $10^{-12}$, which is sufficient to be considered as "perfect coronagraph."

The suggested concept introduces chromatic effect into initially achromatic scheme. Nevertheless only if the bandwidth increases $\sim 30\%$ the performance of a double scheme becomes worse than performance of a one stage coronagraph. For 10% bandwidth the gain in contrast is about 1 order of magnitude.

The contrast can be estimated by means of the contrast curve, connecting the maxima of the residual *PSF*. The approximate analytical expressions for the contrast curve allow avoiding the time consuming numerical simulation. More complex effects, such as chromaticity, can be calculated using knowledge about the *PSF* for segmented mirrors.

## ACKNOWLEDGMENTS

The presented study is accomplished in a framework of the European Extremely Large Telescope Design Study (Framework 6).

## REFERENCES

1. I. Hook "Highlights from the science case for a 50- to 100-m ELT", in *Ground-Based telescopes*, J.Oschmann, ed., Proc. SPIE 5489, pp. 35-46 (2004).

2. M. Ollivier "Experimental aspects of DARWIN", in Astronomy with High Contrast Imaging II, C. Aime and R. Soummer, eds. EAS Publications Series, 12 pp. 235-244 (2004)

3. V. G. Ford, A. B. Hull, S. B. Shaklan, M. B. Levine, M. L. White, A.E.Lowman, E. J. Cohen, "Terrestrial Planet Finder coronagraph", in Techniques and Instrumentation for Detection of Exoplanets, D. Coulter ed., Proc. SPIE 5170, pp. 1-12 (2003).

4. D. Mouillet, T. Fusco, A.-M. Lagrange and J.-L. Beuzit, "Planet Finder on the VLT: context, goals and critical specifications for adaptive optics", in *Astronomy with High Contrast Imaging*, C. Aime and R. Soummer, eds. EAS Publications Series, 8 pp. 193-199 (2004)

5. A. Boccaletti, P. Riaud, P. Baudoz, J. Baudrand, J.-M. Reess and D. Rouan "Coronagraphy with JWST in the thermal IR", in *Astronomy with High Contrast Imaging II*, C. Aime and R. Soummer, eds. EAS Publications Series, 12 pp. 195-204 (2004)

6. P. Dierickx, E. Brunetto, F. Comeron, R. Gilmozzi, F. Gonte, F. Koch, M. LeLouarn, G. Monnet, J. Spyromilio, I. Surdej, C. Verinaud, N. Yaitskova "OWL phase A status report", in *Ground-Based telescopes*, J. Oschmann, ed., Proc. SPIE 5489, pp.391-406 (2004).

7. J. Nelson, "Progress of the Californian Extremely Large Telescope (CELT)", in *Future Giant telescopes,* J.R. Angel, R. Gilmozzi, eds. Proc. SPIE 4840, p. 47- 59 (2002)

8. B. Lyot, MNRAS (1939)

9. M. Kuchner & W. Traub, ApJ, 570, 900, astro-ph/0203455, (2002)

10. A.Shivaramakrishnan, N.Yaitskova "Lyot coronagraphy on Giant segmented-mirror telescopes", The Astrophysical Journal, 626, June (2005)

11. N.Yaitskova, K.Dohlen, P.Dierickx, "Analytical study of diffraction effects in extremely large segmented telescopes" JOSA A/Vol. 20, No. 8, pp. 1563-1575 (2003)

12. L. Jolissaint and J.-F. Lavigne "An analytic model for the study of the impact of mirror segmentation on AO performance and application to a 30-m segmented telescope", SPIE 5497, pp. 349-360 (2004).